# A simple "Boxed Molecular Kinetics" approach to accelerate rare events in the stochastic kinetic master equation.

*Robin Shannon,*[1,2] *David R. Glowacki,*[1,2,3]
[1]Mechanical Engineering, Stanford University, Stanford, CA 94305, USA
[2]School of Chemistry, University of Bristol, Bristol, BS8 1TS, UK
[3]Department of Computer Science, University of Bristol, BS8 1UB, UK

ABSTRACT The chemical master equation is a powerful theoretical tool for analysing the kinetics of complex multi-well potential energy surfaces in a wide range of different domains of chemical kinetics spanning combustion, atmospheric chemistry, gas-surface chemistry, solution phase chemistry, and biochemistry. There are two well-established methodologies for solving the chemical master equation: a stochastic "kinetic Monte Carlo" approach and a matrix-based approach. In principle, the results yielded by both approaches are identical; the decision of which approach is better suited to a particular study depends on the details of the specific system under investigation. In this article, we present a rigorous method for accelerating stochastic approaches by several orders of magnitude, along with a method for unbiasing the accelerated results to recover the "true" value. The approach we take in this paper is inspired by the so-called "boxed molecular dynamics" (BXD) method, which has previously only been applied to accelerate rare events in molecular dynamics simulations. Here we extend BXD to design a simple algorithmic strategy for accelerating rare events in stochastic kinetic simulations. Tests on a number of systems show that the results obtained using the BXD rare event strategy are in good agreement with unbiased results. To carry out these tests, we have implemented a kinetic Monte Carlo approach in MESMER, which is a cross-platform, open-source, and freely available master equation solver.

## 1. Introduction

Predicting the rate of chemical transformations is of fundamental importance to an array of scientific endeavours from atmospheric[1] and combustion chemistry modelling[2] to protein folding[3] and drug binding.[4] In each of these domains, chemical change relies on complex networks of chemical reactions; interrogating these networks or predicting the relative rate coefficients is a very active area of research. Over the years, transition state theory (TST) has proven an invaluable tool for predicting the rate coefficients of chemical reactions. [5, 6] Building upon this, chemical master equation models[7-9] (and analogous Markov state models[10]) have risen to prominence when dealing with complex reacting systems. The specific focus of this paper is the energy grained master equation (EGME), which provides a detailed microcanonical kinetic description of coupled chemical reactions, enabling one to treat the competition between chemical reactions and energy transfer A range of recent studies highlight the prevalence of non-thermal (non-equilibrium) effects, spanning the gas phase, [11, 12] solution phase [13, 14] and the gas surface interface.[15] In all of these systems, the EGME offers a useful tool for first principles kinetic modelling. In this paper we present a new rare event methodology wherein EGME simulations can be accelerated by orders of magnitude.

There are two main strategies used to solve the EGME: kinetic Monte Carlo (KMC) approaches [16, 17] and matrix aproaches.[18, 19] This paper predominantly deals with KMC approaches. The KMC approach to solving the EGME involves running a range of Monte Carlo 'trajectories', each of which has a pre-specified timestep, until statistical convergence is achieved. In this respect, the KMC approach has similarities with molecular dynamics simulations. A consequence of the KMC approach to solving the EGME is the fact that KMC simulations (like MD simulations) are particularly susceptible to the so-called "rare event problem" encountered in the field of molecular dynamics: the problem arises from the fact that the fundamental timescale of each simulation step is often many orders of magnitude shorter than the timescale on which chemical change (or chemical reactions) occur. As a result, the typical waiting time required to observe a chemical event often requires a prohibitive number of simulation steps. Many approaches are used within the molecular dynamics community to alleviate or circumvent the rare event problem, including milestoning,[20, 21] forward flux sampling, [22, 23] metadynamics,[24] umbrella sampling,[25] and Boxed Molecular Dynamics (BXD).[26-30]

In this paper, we set out to investigate whether rare event strategies of the sort which typically find application in molecular dynamics may be adopted to accelerate KMC EGME simulations. Inspired by the BXD approach, this paper outlines a rare event acceleration strategy for use in accelerating KMC solutions to the EGME. We refer to the method as "boxed molecular kinetics" (BXK) and demonstrate its ability to reduce the cost of KMC simulations by many orders of magnitude. In order to demonstrate the acceleration afforded by the BXK algorithm, we have implemented a KMC algorithm in the EGME code MESMER.[19] This is, to our knowledge, amongst the first examples of a code in which both matrix approaches and KMC methods are available in the same framework. As such, it offers an ideal opportunity for the chemical kinetics community to better evaluate the relative merits of each approach in a wide range of chemically important systems.

This paper is organized as follows: Section 2 details the underlying theory of the EGME and its current implementation within MESMER. Section 2a introduces the general form of the EGME and in sections 2b and 2c, the matrix and KMC methods are described in turn. Section 2c then describes the KMC implementation in MESMER and compares matrix method and KMC simulations for a simple, atmospheric system involving the reaction between the acetyl radical and O2. Section 3 describes the background to the BXD methodology and introduces the new BXK approach. The details of the BXD and BXK methods are described in sections 3a and 3b respectively and then in section 3c the BXK approach is used to treat the cyclopentane isomerisation system, which forms part of the MESMER QA test set. Finally section 4 presents some conclusions.

## 2. Master Equation Methodology and Implementation in MESMER

### *2a. EGME Background*

The detailed theoretical basis for the EGME has been described in several publications;[31, 32] therefore we provide only a brief description here. During some time window $dt$, a chemical species or intermediate may do one of two things. It may undergo a chemical transformation (i.e., reaction) into some other intermediate, or it may undergo energy transfer, typically through interaction with a system bath. The chemical master equation comprises a differential rate equation which models the competition between these two possibilities, describing the population $p$ of a particular isomer $m$ with some rovibrational energy $E$ at some time $t$ as follows:

$$\frac{dp_m(E)}{dt} = \omega \int_{E_{m0}}^{\infty} P(E|E')\, p_m(E')dE' - \omega p_m(E) - \sum_{n \neq m}^{M} k_{nm}(E)\, p_m(E) + \sum_{m \neq n}^{M} k_{mn}(E)\, p_n(E)$$

$$- k_{Pm}(E)p_m(E) - k_{Rm}(E)p_m(E) + k_{Rm}(E)K_{Rm}^{eq}\frac{\rho_m(E)e^{-\beta E}}{Q_m(\beta)} p_R N_a$$

(Eq.1)

The first term in Eq. 1 represents the probability that $p_m(E)$ is populated by collisional energy transfer via bath collisions. $\omega$ is the Leonard-Jones collision frequency and $P(E|E')$ is the probability that collision with bath will result in a transition from $p_m(E')$ to $p_m(E)$. The second term represents the loss from $E$ via energy transfer. The product of $\omega$ and $p_m$ effectively gives a collisional energy transfer rate coefficient. The third term represents the loss from $p_m(E)$ via reaction to give other isomers, denoted by subscript $n$. $k_{nm}(E)$ is the microcanonical rate constant for loss from isomer $m$ to isomer $n$. The fourth term represents the increase in $p_m(E)$ by reactions from isomer $n$ that give isomer $m$. The fifth term represents the rate of irreversible loss from $p_m(E)$ to some product, with $k_{Pm}(E)$, representing the corresponding rate of loss. The final two terms are associated with the so-called bimolecular source term. These only apply to those wells that are populated via bimolecular association. Assuming that the bimolecular reactants are thermalized (i.e., exhibit a Boltzmann distribution), and that a pseudo-first order approximation is appropriate, then the sixth term and seventh term represent the respective rates of loss from $p_m(E)$ via re-dissociation to reactants and the rate at which two reactants associate to populate $p_m(E)$. $k_{Rm}(E)$ represents the rate constant for $p_m(E)$ re-dissociation to give deficient reactant $R$, and $K_{Rm}^{eq}$ is the equilibrium constant between isomer $m$ and the bimolecular reactants. $Q_m(\beta) = \sum_E p_m(E)e^{-\beta E}$ is the rovibrational partition function for the molecular species corresponding to isomer $m$, $n_R$ is the population of the deficient bimolecular species and $N_a$ is the concentration of the excess reactant.

When a system includes a bimolecular source term an additional differential equation is required in order to fully define the system:

$$\frac{dp_R}{dt} = \sum_{m=1}^{M} \int_{E_{i0}}^{\infty} k_{Rm}(E) p_m(E) dE -$$

$$p_R N_a \sum_{m=1}^{M} K_{Rm}^{eq} \int_{E_{i0}}^{\infty} k_{Rm} \frac{\rho_m(E) e^{-\beta E}}{Q_m(\beta)} dE$$

(Eq.2)

Eq.2 shows the most typical form for a bimolecular source term, but recent work has shown that is possible to fully generalise the treatment of bimolecular processes in a master equation description such that the assumption that reactant R is thermalized is no longer necessary.[33] It is thus possible to treat energy transfer in R so as to fully satisfy detailed balance. Eqs 1 and 2 give the 'continuum' representation of the master equation; however, it is not generally possible to solve these equations analytically owing to the fact that the terms which depend on energy $E$ tend to have complicated functional forms. In practice Eq 1 and 2 are solved by discretization into 'grains' in energy space, resulting in a set of coupled differential equations. The 'graining' procedure effectively amounts to lumping the energy space of a chemical species into energy grains of a set size. Grain to grain transitions are then defined such that the processes described in the right hand terms of Eq. 1 are captured. Reactive rate coefficients between grains are typically described using Rice Ramsperger Kassel and Marcus (RRKM)[34] theory, and energy transfer between grains of the same species are described using any of a wide range of functional forms which describe energy transfer. A schematic of these processes is shown in Figure 1 and this figure will be discussed in more detail in the next section. Using freely available software packages such as MESMER[19] and Multiwell,[17] it is easy to apply the master equation in order to gain insights into complicated multistep chemical reactions.

As described in the introduction, there are two methods commonly used to solve the EGME: linear algebra matrix aproaches[7-9, 16, 17] and kinetic Monte Carlo (KMC)[16, 17] approaches based upon the Gillespie algorithm.[35] The former rely on casting Eqs 1 and 2 as an eigenvalue problem and carrying out a matrix diagonalization. The latter rely on running a number of time-dependent Monte Carlo simulations constrained to satisfy the transition probabilities in Eqs 1 and 2, and averaging the results over a wide range of initial conditions until some specified level of convergence has been reached. Either approach yields a solution to the EGME in the form of time dependent species profiles and the results one obtains are independent of the method used to solve the EGME – so long as the matrix diagonalization procedure is numerically stable, and so long as enough stochastic trials have been conducted to satisfy a reasonable convergence criterion.

*2b. The Matrix Method*

The matrix approach for solution of the EGME was first used in 1983 by Scranz and Nordholm[36] and has subsequently been extensively developed by a number of groups. [7-9, 16, 17] In this approach the coupled differential equations described by the discretised form of Eq. 1 are formulated as the following eigenvalue problem:

$$\frac{d}{dt}\boldsymbol{n} = \boldsymbol{Mn} \quad \text{(Eq.3)}$$

where $\boldsymbol{n}$ is a vector containing the grain populations of every species and $\boldsymbol{M}$ is a matrix whose elements comprise rate coefficients for the grain to grain energy transfer and reactive processes. In order to solve the master equation, the matrix $\boldsymbol{M}$ must be diagonalised; to exploit efficient diagonalization routines which operate on symmetric matrices, MESMER utilizes detailed balance to symmetrise $\boldsymbol{M}$, forming a symmetric matrix $\boldsymbol{S}$ as follows:

$$\boldsymbol{S}_{ij} = \boldsymbol{M}_{ij}\left(\frac{f_i}{f_j}\right)^{1/2} = \boldsymbol{M}_{ji}\left(\frac{f_j}{f_i}\right)^{1/2} = \boldsymbol{S}_{ji} \quad \text{(Eq.4)}$$

Where $f_i$ is the equilibrium population of grain $i$. Within the matrix approach, solutions of Eq. 3 take the form:

$$\boldsymbol{n}(t) = \boldsymbol{U}e^{\Lambda t}\boldsymbol{U}^{-1}\boldsymbol{n}(0) \quad \text{(Eq.5)}$$

where $\boldsymbol{n}(0)$ contains the initial conditions for each grain at time zero (i.e., $n_i(E_i, 0)$), $\boldsymbol{U}$ is matrix of eigenvectors obtained from diagonalization of $\boldsymbol{M}$, and $\boldsymbol{\Lambda}$ is a diagonal matrix of the corresponding eigenvalues.

In order to illustrate the graining process and the makeup of $\boldsymbol{n}$ and $\boldsymbol{M}$, a schematic is shown in Figure 1 for a simple fictitious system involving two wells, (each with two energy grains) and a single bimolecular source term (A + B ↔ C ↔ D). This figure not only provides insight into the matrix formulation of the EGME; it will also help to understand the implementation of the KMC algorithm within MESMER described in section 2d. In Figure 1 the $k_{ij}$ in $\boldsymbol{M}$ simply represent an effective rate coefficient for transition between grains $i$ and $j$ regardless of whether the transition is due to reaction or energy transfer. The Fig 1 schematic neglects the details of how the effective rate coefficients are obtained. The matrix $\boldsymbol{M}$ can be divided into different blocks and these are colour coded to identify the different regions. The lead diagonal marked in green represents the total loss from each grain $g$ or source term $S$; the first row and column (blue) give transitions between the source term and the grains of the other isomers in the system (in this case between $s_1$ and grains $g_2$ and $g_3$ in species C), the blocks marked orange deal with energy transfer between the different grains of C and D respectively, and the purple blocks correspond to reactive transitions between C and D. It should be noted that in the absence of tunnelling $k_{24}$ and $k_{42}$ would be 0 since these grains span energy regimes entirely below the reaction threshold connecting C and D. In practice, typical EGME simulations will have many more grains than shown here and the matrix $\boldsymbol{M}$ will be significantly more complex, however this figure serves to illustrate the key features of the matrix representation of Eq.1 in its discretised form.

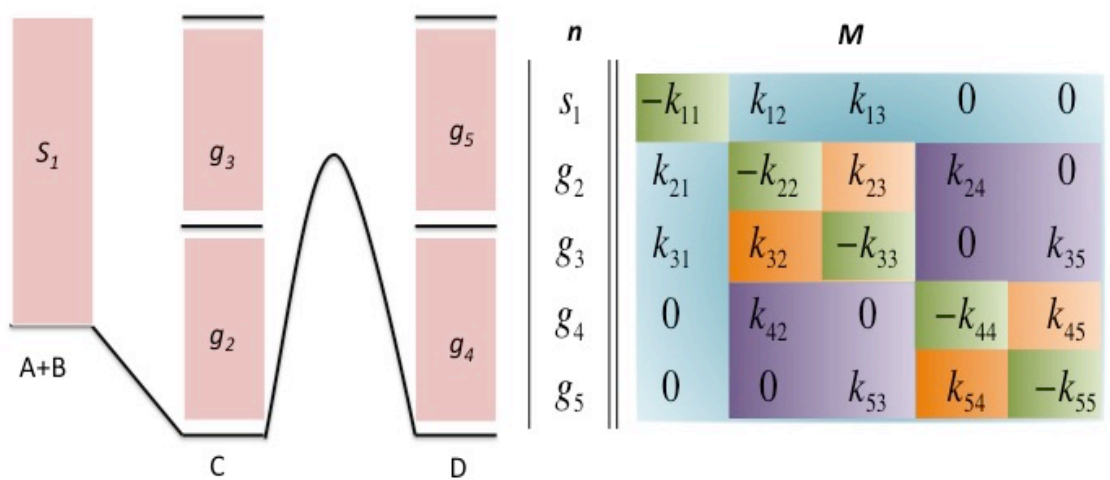

Figure 1: Schematic showing the grained potential energy surface for a fictitious chemical system A + B ↔ C ↔ D and the corresponding vector and matrix $n$ and $M$ from Eq.3. The different blocks of the matrix $M$ are described in the text. The colour coding is described in the main text.

Solution of Eq. 3 gives time dependent grain populations; however, it is often desirable to have a more coarse grained representation of this microcanonical detail – e.g., into phenomenological rate coefficients. In the matrix approach, defining these phenomenological rate coefficients is facilitated using methods related to the work of Bartis and Widom.[37] Broadly, these Bartis Widom methods exploit timescale separations in the eigenvalue spectrum and lump the associated eigenvectors so as to uniquely define macroscopic rate coefficients between all pairs of species in the system. So long as there is a good time separation in the eigenvalue spectrum, then it is often possible to calculate a course-grained kinetic representation which exactly reproduces the time dependent microcanonical grain populations.[38-41]

## 2c. The Kinetic Monte Carlo Method

KMC approaches to solving the master equation typically employ Gillespie's Stochastic Algorithm,[35, 42] which has been successfully applied across many domains.[42-44] Within the context of the EGME, Gilliespie's Stochastic Algorithm has been most commonly used in the form implemented in the Multiwell suite of programs.[17] This algorithm works as follows: First a starting species and its starting energy grain must be defined. The starting grain can either be assigned directly or sampled randomly from some distribution. From this grain, a list of possible transitions to other grains is formed (as outlined in Eq. 1) and the rate coefficients associated with each of these transitions is evaluated. Two uniform random numbers between 0 and 1 are then chosen ($r_1$ and $r_2$): $r_1$ identifies amongst the candidate list of possible transitions, which will occur, and $r_2$ defines the time associated with this transition. The transition, which occurs, is then defined as follows:

$$\sum_{j=1}^{n-1} A_j < r_1 A_T \leq \sum_{j=n}^{l} A_j \qquad (Eq.6)$$

where $A_j$ is an ordered list of the $l$ possible rate coefficients. In this case the transition associated with rate coefficient $A_n$ is picked. The time associated with this transition is then given by:

$$A_T = \sum_{j=0}^{l} A_j \qquad (Eq.7)$$

The time $\tau$ associated with this transitions is assigned as:

$$\tau = \frac{-\ln(r_2)}{A_T} \qquad (Eq.8)$$

After this stochastic step, a new grain is populated, the time $\tau$ in incremented, and the procedure is repeated using an updated rate coefficient list $A_j$ corresponding to the new grain. The total kinetic time is tracked and stochastic steps are performed until some maximum time of interest is reached. This entire procedure is repeated several times (usually beginning from different initial conditions) and the results are averaged until the desired degree of convergence is achieved in the time dependent grain populations.

Another KMC approach to solving the EGME, which has been developed by Vereecken and co-workers, is employed URESAM software.[16] Rather than simulate a stochastic trajectory of a single molecule many times, this approach maps the stochastic trajectory of an ensemble of molecules, but does so only once. Alongside this approach methods have been developed to determine steady state product (DCPD method) and intermediate (CSSPI methods) yields.

Unlike matrix approaches to solving the master equation, KMC methods offer powerful approaches for solving the EGME; however, coarse-graining the results which they produce in order to extract a unique set of phenomenological rate coefficients is not entirely straightforward. Instead, rate coefficients are typically obtained from this sort of simulation through fitting the species profiles with appropriate functional forms.

### *2c. MESMER Kinetic Monte Carlo Implementation*

In what follows, we refer exclusively to the Gillespie KMC approach implemented in Multiwell.[17] The main reason for this is that our BXD-inspired acceleration method is more readily applied to this version of the Stochastic Algorithm as opposed to that used in the URESAM software. Within MESMER, the initial energy of a given stochastic trajectory can either be user-specified or randomly sampled from the initial distribution of a user specified starting species. With a starting grain specified, all the information required to calculate a KMC stochastic trajectory is then contained within the transition matrix $M$. As shown in Figure 1, this matrix contains all the rate coefficients governing reactive and non-reactive processes between grains. Because the KMC implementation does not require diagonalization of M, we use the unsymmetrised matrix $M$ rather than the symmetrized form $S$.

Utilising $M$ and given a starting grain $i$, $A_T$ in Eq.7 is equivalent to $-M_{ii}$ and the $A_j$ can be obtained from the elements of row $i$, in $M$. Specifically the $M_{ij}$ are ordered from smallest to largest and Eq. 6 and Eq. 8 become:

$$\sum_{j=1}^{n-1} M_{ij} < -r_1 M_{ii} \leq \sum_{j=n}^{l} M_{ij} \tag{Eq. 9}$$

$$\tau = \frac{\ln(r_2)}{-M_{ii}} \tag{Eq. 10}$$

The KMC stochastic trajectory evolves until a specified maximum time or termination criteria is reached. Along the way, time-dependent snapshots are stored in order to map the time evolution of the grain population onto a set timescales. A number of stochastic trajectories are performed and the resulting species profiles are averaged. These can be compared directly with the species profile obtained from matrix simulations of the same system. The number of simulations required for convergence will depend upon the system under consideration. Usefully however, the stochastic error is well defined owing to the fact that species populations from a KMC trajectory follow a multinomial distribution.[45] As such the 1σ stochastic error for a species population at any given time is given by the square root of the variance

$$\sigma = \sqrt{\frac{1}{N}f(1-f)} \quad \quad \quad \text{(Eq. 11)}$$

where $N$ is the number of stochastic trials and $f$ is the fractional population of the species of interest. The tolerable lower error limit to the number of stochastic trials can vary significantly depending on the system under investigation and the particular reaction channels of interest.

To the best of our knowledge, the MESMER KMC implementation is amongst the first codes where a bimolecular source term has been incorporated into a KMC EGME. Bimolecular associations are typically avoided in KMC simulations; instead an appropriate "activated distribution" is initialised in the association adduct.[46] Since all rate coefficients in the current implementation are obtained directly from the transition matrix *M*, the KMC implementation in MESMER implicitly includes the option to include bimolecular source terms as well as the more general bimolecular methods that have been implemented in recent years.[33] This enables bimolecular systems to be treated in a genuinely equivalent fashion across both matrix approach and KMC simulations since the same transition matrix M is used in both cases.

### *2d. Testing the MESMER KMC Implementation*

To test the stochastic master equation solver implemented in MESMER, we examined a model system based upon the acetyl + $O_2$ reaction system, which is part of the MESMER test suite and has been studied extensively by several groups using both matrix and KMC EGME's.[47,48] This system is important in atmospheric oxidation chemistry, particularly because the concentration of the acetyl radical impacts the kinetics of peroxy acetyl nitrate (PAN) formation; PAN's importance arises from the fact that it provides a "reservoir" which is capable of transporting $NO_2$ over long distances. Association of the acetyl radical ($CH_3CO$) with $O_2$ molecule leads to the formation of acetyl-peroxy radical (R1), which can then react further through isomerisation (R2) or dissociation processes. The main channels under the conditions of interest here are:

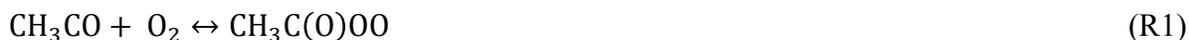
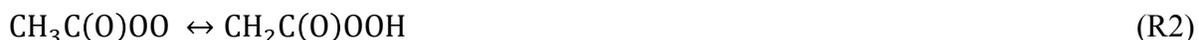

$$CH_3CO + O_2 \leftrightarrow CH_3C(O)OO \quad (R1)$$
$$CH_3C(O)OO \leftrightarrow CH_2C(O)OOH \quad (R2)$$

The system considered here is a model system for the purpose of comparison (the full reaction sequence is somewhat more complex than that given by R1 and R2). Since this system being used as a model system for comparing the two master equation solution methods, the energies of the different species have been altered compared to previous theoretical studies[48] (see Figure 3). Master equation calculations were performed for this system at 298 K and 50 Torr using both the MM and KMC approaches. A grain size of 50 cm$^{-1}$ was used, and a rigid rotor harmonic oscillator approximation was used in calculating the rovibrational state density of each species. Energy transfer was modelled assuming an exponential down model with an average energy transferred upon collision, $<\Delta E_{down}>$, value of 250 cm$^{-1}$ for both $CH_3C(O)OO$ and $CH_2C(O)OOH$. All vibrational frequencies are taken from the study by Carr *et al*[48] and a schematic potential energy is shown in Figure 2. The MESMER input file is given in the supplementary information. The KMC code is not part of the main MESMER distribution and is instead found in the separate KMC branch found at https://sourceforge.net/p/mesmer/code/HEAD/tree/branches/KMC/. Any queries about the KMC version should be directed to the authors of the present work rather than the MESMER forum.

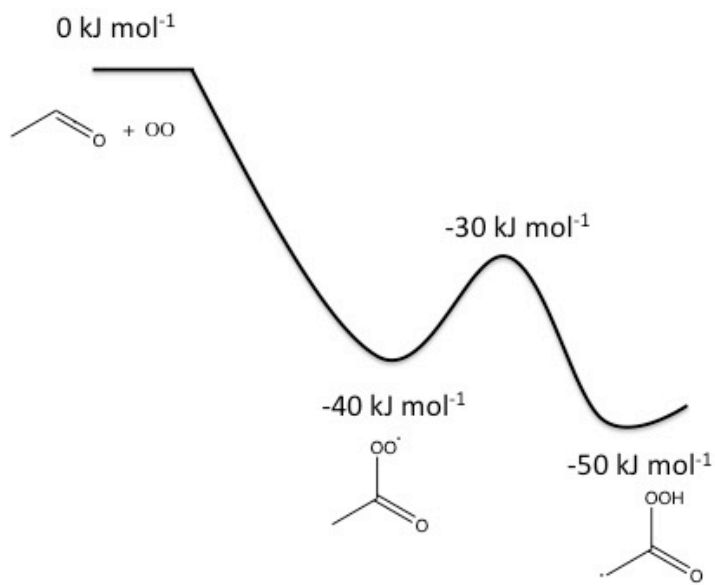

Figure 2: Schematic potential energy surface for a model system based upon the acetyl + O2 reaction.

Figure 3 shows the species profiles obtained for the acetyl radical and also for $CH_2C(O)OOH$ using both methods, with the stochastic simulations averaged over 1000 trials. The decision regarding the number of trials is somewhat arbitrary: we chose to run 1000 trials as a compromise between the size of the stochastic error bars and computational efficiency. In both cases the simulations were initialised with a bimolecular source term (to the best of our knowledge, the first case of a bimolecular source term used with a KMC simulation). Fig 3 clearly shows that the time profiles for the population of acetyl and $CH_2C(O)OOH$ from the matrix method are within the stochastic error bars of the KMC method for the entire simulated timescale.

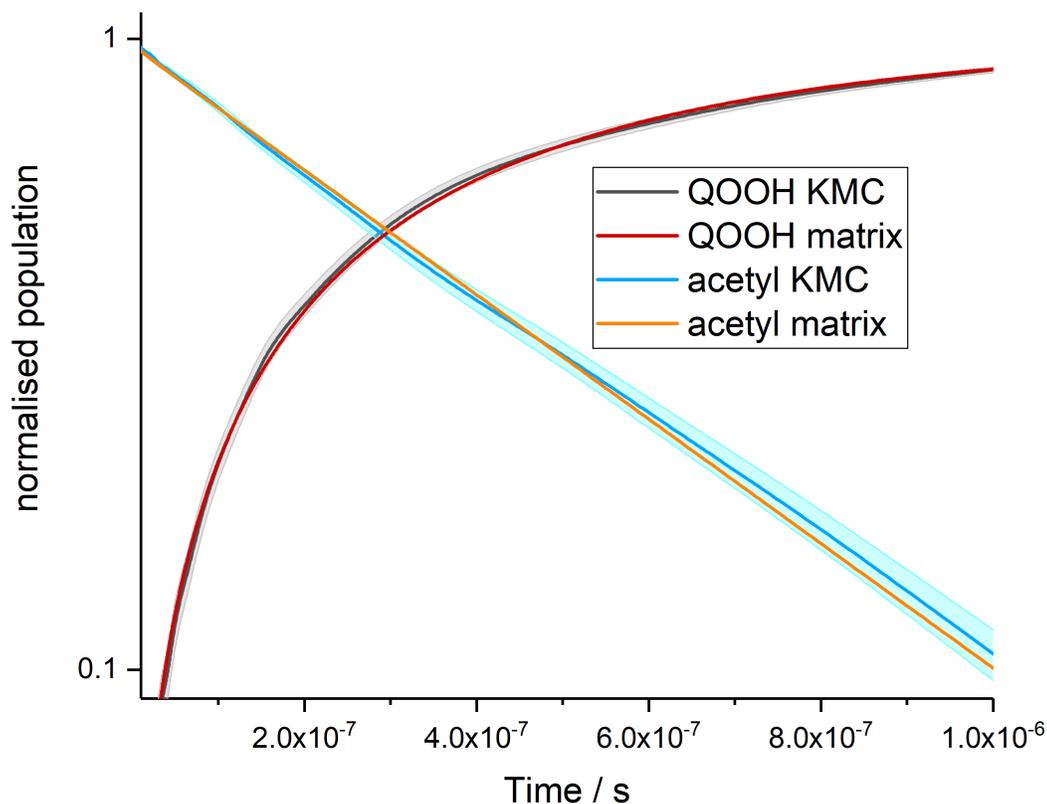

Figure 3: Normalised species populations as a function of time for the acetyl radical and $CH_2C(O)OOH$ as calculated from master equation simulations using a matrix approach and a KMC approach. These simulations were performed at 298 K and 50 Torr. All error bars correspond to the $1\sigma$ stochastic error defined in Eq. 11.

## 3. The BXK Rare Event Acceleration Algorithms.

### 3a. Background

The timestep in a stochastic algorithm is controlled by the inverse of the fastest rate coefficient(s). If there is a large separation of timescales between processes *a* and *b* (i.e., $k_a \gg k_b$), then a large number of simulations may be necessary to observe a statistically significant sampling of rare events. The case of thermal unimolecular reaction over a large barrier provides a good case in point: the time scales of energy transfer processes are separated from the time scales for reactive process by many orders of magnitude. There have been a number of approaches used to address this problem, including tau leaping[49, 50] as well as multiscale steady state and partial equilibrium approaches.[51, 52] These approaches exploit timescale separations in a reactive system and have been successfully employed in many implementations of Gillespie's algorithm. However such approaches have not yet been utilised in the context of the EGME and are not obviously amenable to the type of problem described here. Vereecken and co-workers[16] have developed the DCPD and CSSPI methods for obtaining long-time steady state distributions of rare products and intermediates for a given EGME, but these methods do not give the time dependent species populations.

In fact, the type of rare event problem which one faces running stochastic simulation algorithms is very similar to the rare event problem that occurs in the field of molecular dynamics simulations. This recognition inspired us to apply acceleration methods utilised by the MD community to the stochastic master equation. Specifically, the last few years have given rise to a class of sampling methods in which molecular configuration space is divided into a set of boundaries (also called interfaces or hypersurfaces), and short trajectories are run between boundaries. These methods include milestoning, [20, 21] forward flux sampling, [22, 23] transition interface sampling, [53] nonequilibrium umbrella sampling, [54] and others [55-57]. Owing to the fact that it has been derived as an exact extension of transition state theory (TST), the boxed molecular dynamics BXD[26, 28, 29, 58] method, which we have been actively developing over the last few years (and for which we now have an adaptive implementation) is particularly interesting to consider with respect to the rare event problem one encounters in solving a KMC EGME.

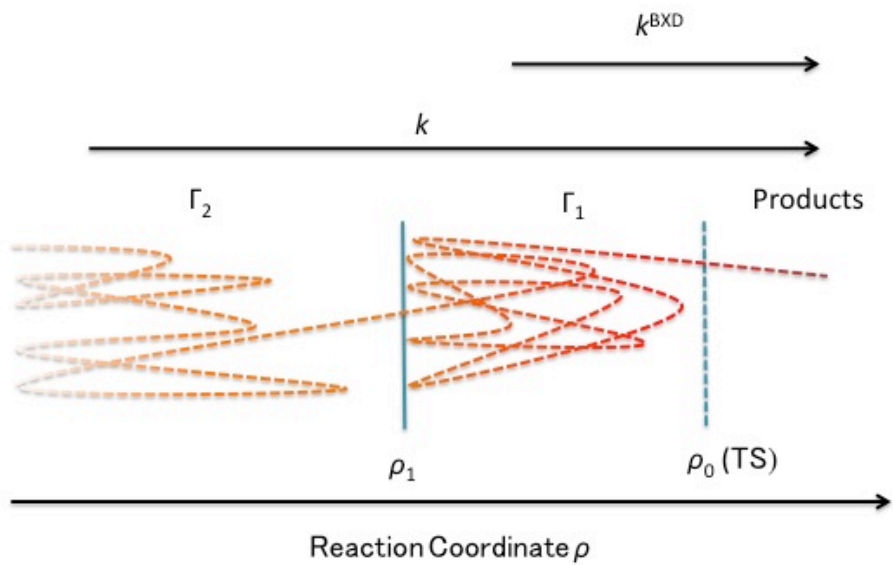

Figure 4: Schematic representation of the original implementation of boxed molecular dynamics BXD. Here a fictitious trajectory penetrates from phase space volume $\Gamma_2$ and $\Gamma_1$ and is confined in this region to accelerate crossing into the product region. The accelerated rate coefficient $k^{BXD}$ can be readily corrected to return $k$ as described in the text.

The simplest implementation of the BXD method is illustrated in Figure 4. According to classical TST, the phase space of the system is separated into reactant and product regions by a dividing surface at $\rho_0$ along some reaction coordinate, and the reaction rate coefficient is then calculated as a flux through the dividing surface. BXD accelerates passage through $\rho_0$ by splitting the reactant phase space into two "boxes": $\Gamma_1$, which spans $\rho_0$ to $\rho_1$, and $\Gamma_2$, which is bounded by $\rho_1$. By locking the dynamics within $\Gamma_1$ using a velocity inversion algorithm, the trajectory crosses the transition state more often, yielding an accelerated rate coefficient, $k^{BXD}$. The actual (unbiased) rate coefficient, $k(T)$, of going from the reactant region, $\Gamma_1 + \Gamma_2$, to the product region, $\Gamma_0$, may then be recovered as:

$$k(T) = k^{BXD} \times P^{CORR} \tag{Eq. 12}$$

where the $P^{CORR}$ correction factor is the probability of finding the system in $\Gamma_1$. Provided the assumption of equilibrium between boxes $\Gamma_1$ and $\Gamma_2$ is valid, $P^{CORR}$ is calculated simply as the fraction of the phase volume $\Gamma_1$ to the total reactant phase volume.

$$P^{CORR} = \frac{\Gamma_1}{\Gamma_1 + \Gamma_2} \qquad (Eq.\ 13)$$

The phase volume ratio in (Eq. 13) may be estimated from a Monte-Carlo random walk or by running a trajectory in $\Gamma_1 + \Gamma_2$. Another way to calculate the correction factor is to recognize that the ratio of the two phase volumes is simply the equilibrium constant $K_{21}$ of exchange between $\Gamma_1$ and $\Gamma_2$, which can be estimated from classical molecular dynamics as a ratio of the box-to-box rate constants $k_{12}$ and $k_{21}$:

$$P^{CORR} = \frac{1}{1+\frac{\Gamma_2}{\Gamma_1}} = \frac{1}{1+K_{21}} = \frac{1}{1+\frac{k_{12}}{k_{21}}} \qquad (Eq.\ 14)$$

The fundamental efficiency gain of BXD derives from the fact that it is less expensive to converge $k^{BXD}$ and $P^{CORR}$ separately than their small product $k(T)$

In a molecular dynamics context, the equilibrium constant between boxes is typically converged through repeated collisions on each side of the reflective boundary. However in a statistical mechanical context like that of the energy grained master equation, we have access to the densities of states of every grain whose dynamics is treated in our transition matrix. As a result, we can calculate this equilibrium constant in a straightforward way, from the Boltzmann populations of the energy grains of the species in question.

### *3b. BXK Algorithm*

The boxed molecular kinetics (BXK) procedure is implemented in MESMER as follows: For a given well, a boundary is placed at a user-specified grain. Those grains with an energy which is lower than the boundary define $\Gamma_2$ and those grains at higher energies define $\Gamma_1$, as shown in

Figure 5. Within the BXK methodology the starting grain will be the lowest grain in $\Gamma_1$. The grains in $\Gamma_2$ are then excluded from the stochastic trajectory and matrix elements $M_{ij}$ referencing grains $j$ which reside in $\Gamma_2$ are ignored.

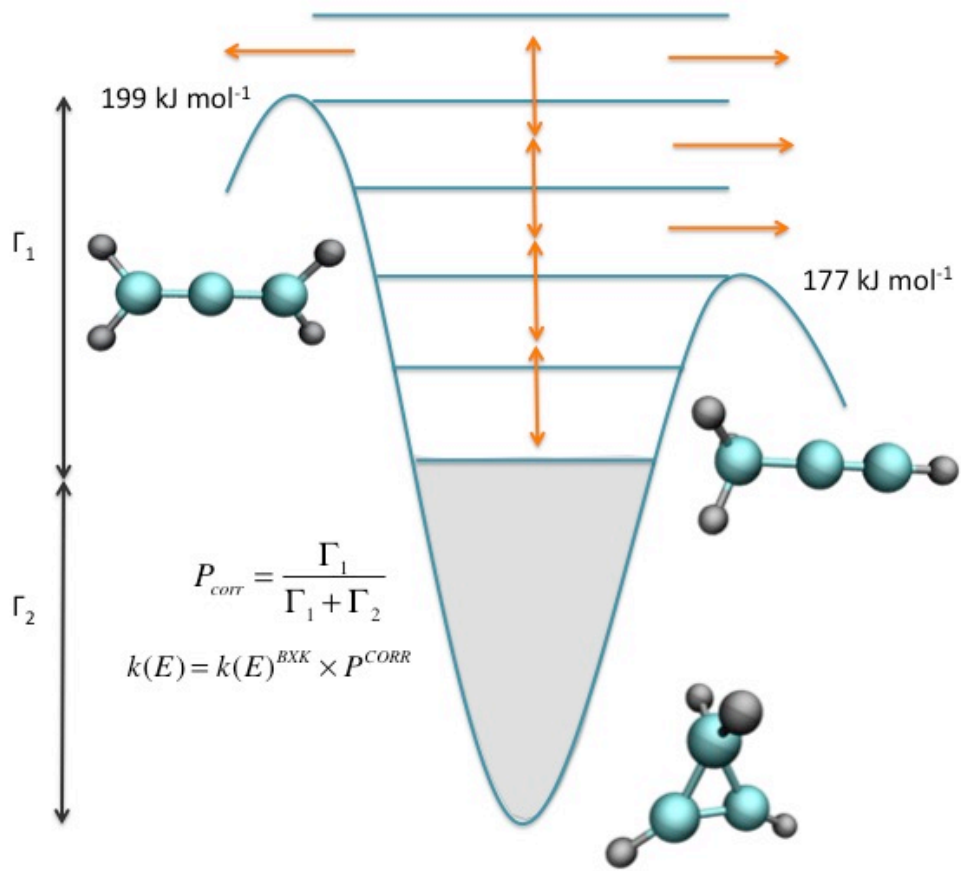

Figure 5: Schematic potential for the cyclopropene isomerisation system demonstrating the application of BXK to a KMC EGME. The grey region shows the region of the cyclopropene energy space, which is excluded by the BXK boundary and the orange arrows represent collisional and reactive transitions available to the constrained EGME. $P^{CORR}$ is the correction factor which is applied to all constrained rate coefficients in order to unbias the results. Barrier energies are taken from Miller and Klippenstein[61] and are given in kJ mol$^{-1}$

If $s$ is the index of the first grain in $\Gamma_1$, a modified rate coefficient list replaces $A^T$:

$$A^{BXK} = \sum_{j=s}^{n} M_{ij} \qquad \text{(Eq. 16)}$$

where $i \geq s$. Eq. 9 and Eq. 10 are slightly modified as follows:

$$\sum_{j=s}^{n-1} M_{ij} < r_2 A^{BXK} \leq \sum_{j=n}^{l} M_{ij} \tag{Eq. 17}$$

$$\tau = \frac{\ln(r_1)}{A^{BXK}} \tag{Eq.18}$$

The altered KMC implementation now only considers grains in $\Gamma_1$. The transition rates for the resulting truncated system (excluding $\Gamma_2$) are then corrected in a typical BXD manner. In this case the standard BXD correction is trivial since the normalised equilibrium population of each grain, $p_i$, is known from the rovibrational density of states giving:

$$P^{CORR} = \frac{\Gamma_1}{\Gamma_1 + \Gamma_2} = \frac{\sum_{j=s}^{l} p_j}{\sum_{i=1}^{l} p_i} \tag{Eq. 19}$$

This correction is performed at the microcanonical level as follows:

$$k(E) = k(E)^{BXD} P^{CORR} \tag{Eq. 20}$$

This BXK correction is applied to all $M_{ij}$ with $i,j >= s$. This new BXK methodology has similarities with reduced matrix[59] and reservoir state[60] approaches which have been previously applied to matrix solutions of the EGME.

### *3c. Application of the BXK Algorithm to Cyclopropene Isomerisation*

To test the BXK approach, simulations were carried out on the cyclopropene isomerisation reaction, which is part of the MESMER test suite. This is an example of a thermal reaction across a large energy barrier and would typically be impossible to treat using a KMC approach at reasonable temperatures due to the rarity of reaction events. Such processes can also be challenging even with matrix methods, and the so-called 'reservoir-state' approximation is often employed.[60, 62] A MESMER input file for this system utilizing the BXK approach is given in the supplementary information. Figure 5 demonstrates the application of a BXK approach to this system, in which cyclopropene can undergo two competitive isomerisation reactions to either allene or propyne. As shown in Figure 5, both channels have energy barriers in excess of 177 kJ

mol$^{-1}$. Calculations were performed for this system at 100 Torr and 800 K with a grain size of 100 cm$^{-1}$. Energy transfer in the cyclopropene was modelled using an exponential down model with an average energy transferred upon collision, $<\Delta E_{down}>$, value of 250 cm$^{-1}$. A harmonic oscillator rigid rotor approximation was assumed when calculating the ro-vibrational density of states; frequencies, rotational constants, and potential energies were taken from a study by Miller and Klippenstein. [61] It should be noted that in this case the two possible isomerisation channels of cyclopropene were considered to be irreversible for simplicity.

It is known that the exact nature of the BXD correction in molecular dynamics relies on the assumption that the dynamics in region $\Gamma_1$ is uncorrelated (loses all memory of its past behaviour) prior to passage through $\rho_0$ and into the product reaction. As such it is necessary for $\rho_0$ and $\rho_1$ to be sufficiently separated such that reaction does not occur instantaneously. Likewise with the BXK method, it would be expected that the simulated EGME species profiles and corresponding rate coefficients converge to the unbiased result as the energy of the BXK boundary $\rho_1$ is lowered relative to the lowest reaction threshold (177kJ mol$^{-1}$ in the current case).

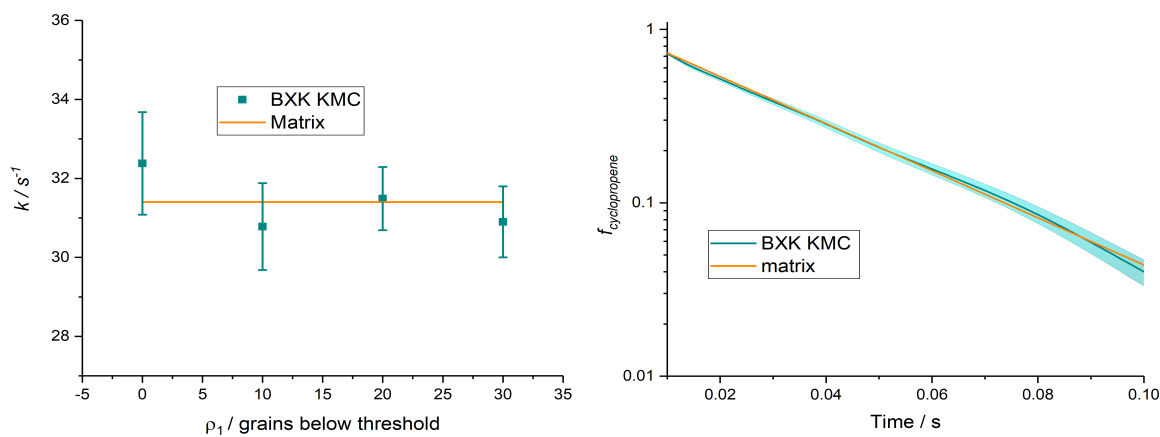

Figure 6: Comparison between BXK KMC simulations and matrix method simulation for the cyclopropene system at 100 Torr and 800 K. The left-hand panel gives unbiased BXK rate coefficients as a function of BXK boundary ($\rho_1$) placement. The rate coefficients for the KMC simulations are obtained through fitting cyclopropene decays to Eq. 21; matrix method rate coefficients are obtained from a Bartis Widom analysis. The left panel shows a direct comparison of normalised cyclopropene concentration profiles obtained from a matrix simulation and a KMC BXK simulation with a $\rho_1$ 10 grains below the lowest transition state

In Figure 6 we explore the rate coefficients obtained from BXK KMC simulations as a function of the BXK boundary ($\rho_1$) energy. Rate coefficients were obtained through fitting the species profiles from the BXK EGME simulations with the exponential decay expression:

$$f_{cyclopropene}(t) = e^{-kt} \tag{Eq. 21}$$

where $f_{cyclopropene}$ is the normalised population of cyclopropene from the species profiles and $k$ is the total loss rate coefficient for cyclopropene. An example exponential fit is given in Figure S1 of the supplementary information. These fits were performed using the Origin graphical analysis software. For comparison, rate coefficients from matrix EGME are shown for the same system as obtained through a Bartis Widom analysis. From Figure 6 it can be observed that the BXK rate coefficients are in agreement with the Bartis Widom rate coefficients (obtained from the matrix method) for a range of values of the BXK boundary placement $\rho_1$ (see Figure 5). The lowest reaction threshold(176.6 kJ mol$^{-1}$ in this case) should be viewed as a hard upper limit to the placement of $\rho_1$. In some cases, particularly at low pressures, it may be necessary to place $\rho_1$ substantially below the reaction threshold in order to converge the BXK simulations. This will be discussed further in the section on practical implementation  In the right hand panel of Figure 6, two species profiles are compared directly without post-analysis to obtain rate coefficients. In this case the BXK species profile corresponds to the simulation with a $\rho_1$ placement 10 grains below the reaction threshold.

| $\rho_1$ / grains below threshold | Average execution time for 1 KMC trajectory / s |
|---|---|
| 0 | 0.66±0.07 |
| 20 | 6.77±0.51 |
| 40 | 118.50±0.90 |
| 60 | 621.43±6.63 |
| 80 | 4800.90±43.31 |
| no BXK* | 1×10$^7$ |

Table 1: Average execution time for 1 BXK KMC trajectory, for different energies of the BXK boundary $\rho_l$ at or below the transition state energy of 176.6 kJ mol$^{-1}$. * The value at a $\rho_l$ energy of 0 is estimated as described in the text above

Table 1 shows how execution time depends on the $\rho_l$ energy. It can be seen that simulations with a $\rho_l$ energy equal to the lowest transition state are over 3 orders of magnitude faster than BXK simulations with the lower $\rho_l$ energy of 96.6 kJ mol$^{-1}$. We were unable to perform unbiased KMC simulations for this system owing the very long time for convergence, but we can estimate the number of stochastic steps required as follows: the species profiles in the left panel of figure 6 demonstrate that in order to observe majority of cyclopropene loss it is necessary to run simulations for 0.1 seconds. For this system, each timestep in the stochastic simulation is on the order of $1\times10^{-10}$ seconds, which would mean that roughly $1\times10^9$ simulation steps would be necessary in order to perform the KMC simulations described above without a BXK correction. On typical single-core desktop architectures, it is possible to perform approximately 100 stochastic trials per second; this suggests an unbiased execution time for a single stochastic trajectory of $1\times10^7$ seconds (~115 days). From this, we estimate that the BXK method accelerates KMC simulations of this cyclopropene system by close to 7 orders of magnitude. All timings are for simulations carried out on a serial 2.6 GHz Intel SandyBridge core running Linux.

*3d. Practical Considerations*

As already noted, the placement of $\rho_l$ at which the BXK simulations converge will depend upon the system and the conditions of interest. We have shown that for the cyclopropene simulation at 1000 Torr and 800 K, $\rho_l$ may be placed at the reaction threshold. This holds true simulations at higher pressure also. At low pressures however $\rho_l$ needs to be placed substantially lower in order to converge the BXK simulations. Figure 7 compares species profiles from BXK simulations for the cyclopropene simulations at 800K and the low pressure of 1 Torr to the matrix method species profile obtained under the same conditions. Here it can be seen that $\rho_l$ needs to be placed at least 50 grains below the barrier in order to converge to the matrix method result. Below this $\rho_l$ placement the BXK simulations are observed to converge within statistical error.

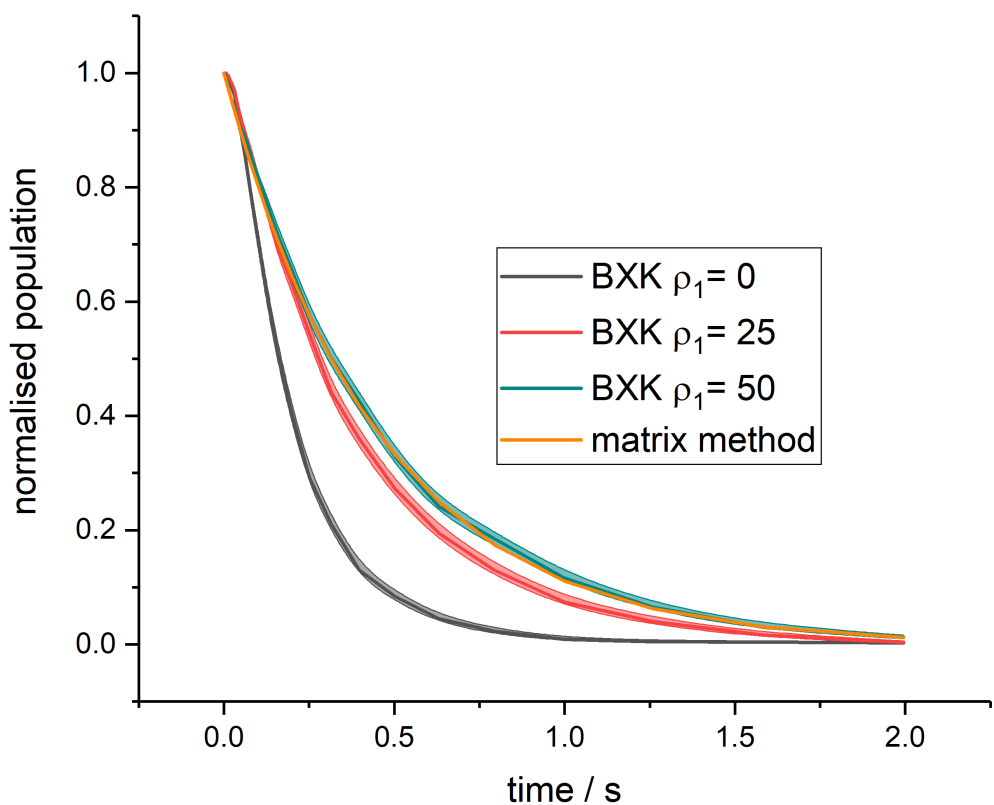

Figure 7: Comparison between BXK KMC simulations and matrix method simulation for the cyclopropene system at 1 Torr and 800 K. The grey red and green lines correspond to cyclopropene populations profiles from simulations with the BXK barrier $\varrho_l$ set at 0, 25 and 50 grains below the reaction threshold. The orange line shows results from matrix method simulations.

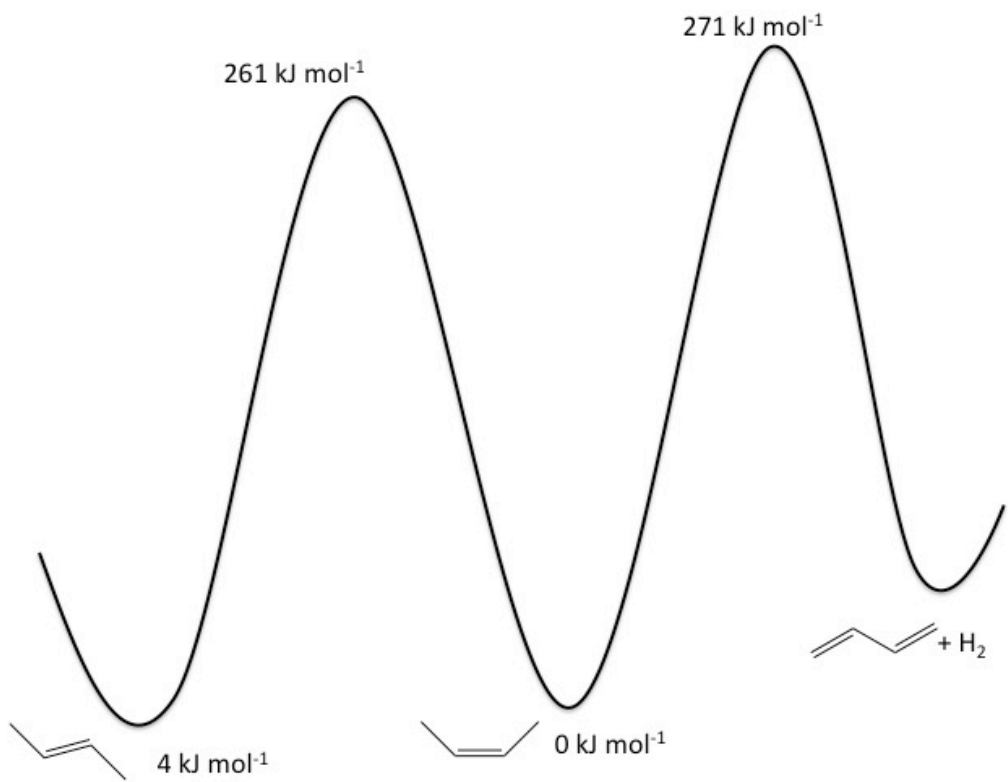

Figure 8: Schematic potential energy surface for the but-2-ene isomerisation / dissociation system.

When dealing with more complex systems there are additional considerations, which are necessary when applying the BXK correction to a given well. By definition the BXK approach is only applicable to a thermal reaction and the correction should not be applied to any wells which may undergo a prompt or activated process. Let us consider the but-2-ene isomerisation and dissociation system, the potential energy surface for which is shown in Figure 8. Assuming we start with a Boltzmann distribution in cis-but-2-ene, the BXK correction may be safely applied. However if the isomerisation to form trans-but-2-ene occurs this new species must be thermalized before the BXK correction can be applied to the new well. To ensure this we introduce an additional parameter, which tracks the number of consecutive collisional events that occur in the stochastic trajectory and only applies the BXK to a new well correction once a specified number of such collisional events have occurred. When applying a BXK approach to a new system, this number should be varied along with the placement of the BXK boundary until the resulting species profiles are observed to converge. Figure 9 shows BXK simulations at 0.003

Torr and 1000 K for the but-2-ene system with a BXK boundary 60 grains below the lowest barrier and a thermalization parameter (number of consecutive collisional event prior to BXK) of 100. These are compared to exact solution from a matrix simulation on the same system and good agreement can be observed. An example input file for this system is given in the supplementary information.

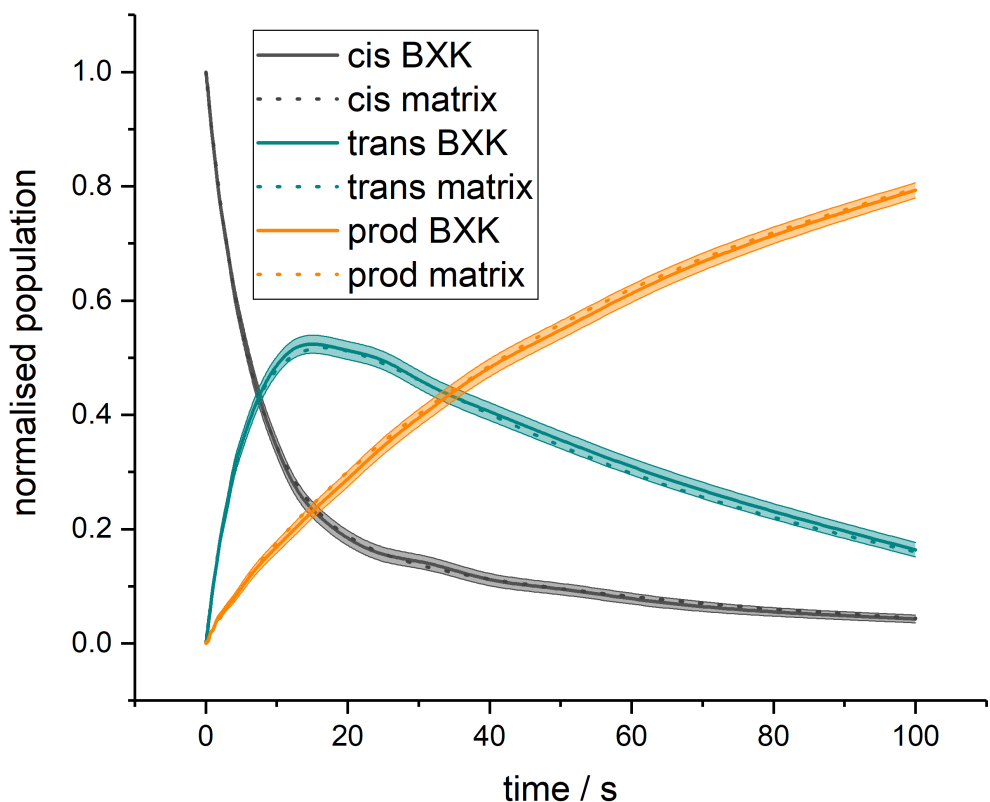

Figure 9: Comparison between BXK KMC simulations and matrix method simulations for the but-2-ene system at 0.003 Torr and 1000 K. Dotted lines are matrix method simulations and solid lines are BXK simulations. The figure shows the time resolved populations of all three species, cis-but-2-ene (grey), trans-but-2-ene (green), and the dissociation products butadiene + $H_2$ (orange).

*4. Conclusions*

In this study we have outlined the BXK acceleration methodology, which has been implemented within the MESMER software package, and has the potential to accelerate rare event sampling in a stochastic trajectory by several orders of magnitude. This greatly expands the range of applications for which a KMC approach to solving the EGME can be used. To our knowledge this is the first instance where both approaches have been coded into the same framework and we demonstrate that both KMC and matrix approaches agree in the limit of infinite stochastic trials. When comparing EGME simulations with molecular dynamics studies, a KMC simulation has certain advantages. For example, a KMC simulation gives information on the energy fluctuations within a molecule which cannot be obtained from an matrix method and this allows for direct comparison with quantities such as energy auto-correlation functions from molecular dynamics results. Recent molecular dynamics studies have demonstrated fascinating non-thermal chemistry in liquid[13, 63-65] and solid phases[15] and the accelerated KMC approach described here could prove useful in bridging the gap between molecular dynamics and matrix master equation calculations for such systems.

AUTHOR INFORMATION

Corresponding Author

*Robin J. Shannon: E-mail: shannon.rj@gmail.com

*David Glowacki: E-mail: glowacki@bristol.ac.uk

Funding Sources

Funding for D. R. G. is from the royal society and EPSRC grant EP/P021123/1. Funding for R. J. S was provided by the US Air Force Office of Scientific Research (AFOSR) under contract no. FA9550-16-1-0051 and researcher mobility grant from the RSC

ACKNOWLEDGMENT

The authors would like to thank Struan Robertson for valuable discussion.